\documentclass[conference]{IEEEtran}
\IEEEoverridecommandlockouts
\usepackage{cite}
\usepackage{amsmath, amssymb, amsfonts}
\usepackage{algorithmic}
\usepackage{graphicx}
\usepackage{textcomp}
\usepackage{xcolor}
\usepackage{listings}
\usepackage{amsmath}

\usepackage{longtable}
\usepackage{hyperref}
\usepackage{caption}
\usepackage{float}
\def\BibTeX{{\rm B\kern-.05em{\sc i\kern-.025em b}\kern-.08em
    T\kern-.1667em\lower.7ex\hbox{E}\kern-.125emX}}
\begin{document}

\title{Boosting Cybersecurity Vulnerability Scanning based on LLM-supported Static Application Security Testing
}

\author{\IEEEauthorblockN{1\textsuperscript{st} Mete Keltek}
\IEEEauthorblockA{\textit{University of Cologne} \\
Cologne, Germany}
\and
\IEEEauthorblockN{2\textsuperscript{nd} Rong Hu}
\IEEEauthorblockA{\textit{Hunan University}\\
Changsha, China \\
upupwords@hnu.edu.cn}
\and
\IEEEauthorblockN{3\textsuperscript{rd} Mohammadreza Fani Sani}
\IEEEauthorblockA{\textit{Microsoft}\\
Copenhagen, Denmark \\
mfanisani@microsoft.com}
\and
\IEEEauthorblockN{4\textsuperscript{th} Ziyue Li$^*$\thanks{$^*$ Corresponding author.}}
\IEEEauthorblockA{\textit{University of Cologne}\\
Cologne, Germany \\
zlibn@wiso.uni-koeln.de}
}

\maketitle

\begin{abstract}
In the fast-evolving landscape of cybersecurity, Large Language Models (LLM's) play a pivotal role, continually improving their ability to analyze software code. This paper introduces a novel approach to vulnerability scanning by integrating conservative SAST (Static Application Security Testing) scanners with LLM capabilities, resulting in the creation of LSAST (LLM-supported Static Application Security Testing). Our approach significantly enhances the performance of LLM's in vulnerability scanning, establishing a new standard in this field. We benchmark LSAST’s efficiency and compare its results with a state-of-the-art LLM. Additionally, we address the inherent drawbacks of LLM's in vulnerability scanning: their reliance on static training datasets, which leads to the exclusion of the latest vulnerabilities, and the privacy concerns associated with sending code to third-party LLM providers. To mitigate these issues, we utilize a open-source LLM to ensure privacy and employ a novel approach to gather relevant vulnerability information, thereby equipping the LLM with up-to-date knowledge.
\end{abstract}


\section{Introduction}
\label{sec:Intro}
The realm of cyber security is characterized by its dynamic nature, evolving rapidly alongside advancements in Large Language Models (LLM's). These sophisticated models continuously enhance their ability to analyze and comprehend software code, due to their access to comprehensive training data, which includes both code and vulnerability information. 

Using large language models (LLM's) for vulnerability detection in code has already shown promising results. In this paper, we aim to enhance the capabilities of LLM's in identifying vulnerabilities in software code. Our approach combines conservative Static Application Security Testing (SAST) scanners, such as Bearer, with LLM's and thereby achieves synergy in code scanning. We accomplish this by engineering a prompt that includes the results of a conservative SAST scanner and instructing the LLM to search only for new vulnerabilities not identified by the SAST scanner. This method allows the LLM's task to be strongly supported by the SAST scanner results, leading to higher performance in identifying additional vulnerabilities. Our findings indicate that this approach significantly enhances the LLM's vulnerability detection capabilities and improves the quality of the results, making this method crucial for future use of LLM's in SAST scanning.

However, a notable limitation of LLM's is their reliance on a static training dataset, leading to the exclusion of the latest vulnerabilities. This creates significant security risks, when utilizing LLM's for Vulnerability Scanning, as vulnerability scanners rely on access to the most current vulnerabilities to adequately address emerging threats.

Furthermore, utilizing LLM's for vulnerability scanning introduces additional challenges, notably when transmitting code to API endpoints of third-party LLM providers. This practice exposes sensitive code, opening the door to potential privacy and security risks, notably unauthorized access by malicious entities. Additionally, there's the concern of the code being repurposed for future model training, creating a vulnerability to training data extraction attacks. Addressing these concerns is necessary to prevent malicious exploitation of possible vulnerabilities in the code.

In this paper, we aim to tackle these challenges head-on. Our approach centers on utilizing a locally hostable instance of Llama-3, to mitigate security and privacy risks associated with third-party LLM providers. Remarkably, existing research has acknowledged these privacy and security issues but has yet to benchmark open-source LLM's for Vulnerability Scans. Therefore, our paper represents the first attempt to showcase the efficiency of a locally hostable LLM-Scanner in identifying vulnerabilities.

To ensure LLM's have access to the most up-to-date vulnerability knowledge, we implement a state-of-the-art Knowledge Retrieval system. This involves aggregating the latest vulnerability reports from HackerOne, a prominent Bug Bounty platform known for its vast repository of reports submitted by ethical hackers worldwide. Leveraging the Retrieval-Augmented Generation Framework, we enable LLM's to acquire knowledge of the most recent vulnerabilities.

By addressing privacy and security concerns, integrating current vulnerability knowledge into LLM's, and enhancing results through SAST scanner outputs, we introduce LSAST (LLM-supported Static Application Security Scanning), heralding a new era in static analysis within IT security. 

Our research further aims to benchmark LSAST's efficiency by comparing its detection capabilities to those of unenhanced LLMs. Additionally, we evaluate LSAST's performance in real-world scenarios by scanning a Python package used in industry. This analysis demonstrates that LSAST can identify vulnerabilities overlooked by conventional scanners, underscoring its importance as an integral component of comprehensive security scanning setups.

\section{Background}
\label{sec:Background}
\subsection{Overview}
\label{subsec:Background/Overview}
With the global cost of cybercrime projected to surge from \$8.15 trillion to \$13.82 trillion \cite{statista2023} and hackers launching an average of 26,000 attacks daily, or one every three seconds \cite{forbes2021}, it is clear that cybersecurity is more critical than ever in the digital age.

As the IT landscape expands, cyber threats have evolved from simple attacks to sophisticated and persistent threats. To counter these threats, a range of cybersecurity measures must be implemented.

One widely adopted measure is security scanning, which plays a crucial role in identifying potential risks within applications or networks.

\subsection{Different kinds of security scanners}
\label{subsec:Background/scaners}
Predominantly, security scanning can be classified into two categories: DAST (Dynamic Application Security Testing) and SAST (Static Application Security Testing).

DAST scanners treat the system or application as a black-box, meaning they operate without knowledge of the internal workings. They identify weaknesses by interacting with the system or application during runtime. By sending specially crafted requests and analyzing the responses, DAST scanners mimic threat actors attempting various types of attacks, such as SQL Injection, Cross-Site Scripting, and others.
SAST scanners, also known as Source Code Analysis Tools, examine the source code without executing the application. They search for specific patterns in the code that are known to be vulnerable.
Both types of scanners complement each other, as they can detect vulnerabilities that the other type might miss. Thereby it is crucial for stakeholders to incorporate both types of scanners into their infrastructure. 

\subsection{Evolution of Large Language Models (LLM's)}
\label{subsec:Background/llm}
The evolution of Large Language Models (LLM's) has significantly advanced their capabilities in vulnerability detection, with considerable potential for further growth. LLM's have rapidly progressed, highlighted by key developments such as OpenAI’s GPT-4 \cite{achiam2023gpt} and Google Gemini 1.5 \cite{team2023gemini}. These models use token prediction to generate human-like text, demonstrating exceptional proficiency in various tasks related to language understanding. Recently, LLM has been applied to more interdisciplinary domains, such as code generation \cite{sui2023reboost,zhang2024benchmarking,li2024pet,yang2024sql}, task planning and tool usage \cite{ruan2023tptu}, LLM agent \cite{kong2024tptu,zhang2024controlling}, and even time-series prediction \cite{liu2024spatial,ye2024survey,liu2024timecma,zhang2024dualtime}, 

\subsection{Integration of LLM's in Cyber Security}
\label{subsec:Background/integration}
Their ability to comprehend and generate text has made it possible to utilize LLM's in software-related tasks. By recognizing and analyzing patterns in vast amounts of data, LLM's have the potential to significantly enhance traditional vulnerability scanning techniques.

The use of LLM's for both SAST (Static Application Security Testing) and DAST (Dynamic Application Security Testing) has proven feasible due to their advanced language processing capabilities. LLM's can identify vulnerabilities that may be missed by traditional pattern matching and fuzzing techniques employed by conventional scanners.

\subsection{Privacy Concerns with LLM's}
\label{subsec:Background/privacy}
Despite these advancements, the use of LLM's raises significant privacy concerns, particularly regarding the handling and sharing of sensitive code data. Organizations must consider the potential risks associated with sending proprietary or confidential code to external LLM providers. These risks include data breaches due to vulnerabilities in network transmission, unauthorized access by LLM providers to sensitive intellectual property, compliance issues with data privacy regulations, and the loss of control over how the code is stored and processed once it leaves the organization's infrastructure.

To mitigate these risks, organizations can use open-source LLM solutions that allow them to host these models locally. By hosting open-source LLM's on their own infrastructure, organizations can reduce the risk of unauthorized access and data breaches while ensuring compliance with data privacy regulations and maintaining greater control over their sensitive data.

\subsection{Need for Up-to-Date Vulnerability Information}
\label{subsec:Background/new-vuln-info}
In addition to privacy concerns when using LLM's for vulnerability detection, there are also issues regarding the absence of up-to-date vulnerability information. LLM's are trained on extensive datasets compiled over a period leading up to the model's creation. These datasets encompass a wide array of text and information from sources such as books, articles, and web content available up to that point. During training, these datasets must remain static to provide a stable sample of language patterns and context. Consequently, a cut-off date is introduced after which the training data is no longer updated, leading to the exclusion of recent events or information from the model's knowledge.

The dynamic nature of the cyber security landscape requires constant awareness of the latest threats. New vulnerabilities, attacks, and coding errors are discovered frequently, and if identified after the cut-off date, they may not be included in the LLM's knowledge base. While LLM's primarily detect logic flaws rather than specific vulnerabilities, new types of attacks can introduce patterns or coding practices that may not be well represented in the training data of the model.

Therefore, relying solely on the static training dataset of the LLM may face limitations in detecting emerging security risks. This underscores the need for integrating up-to-date vulnerability information to enhance the effectiveness of LLM's in vulnerability detection.

\subsection{Retrieval-Augmented Generation}
\label{subsec:Background/rag}

To address this issue, we will introduce a new method of providing up-to-date vulnerability information to the LLM.

We will leverage Retrieval-Augmented Generation, a technique that enhances the capabilities of language models by integrating relevant external knowledge during the answer generation process. This external knowledge is stored as vector representations in databases, allowing it to be queried through similarity searches.

By conducting similarity searches, we can retrieve only the relevant external knowledge from these vector databases. This information is then used to enrich the LLM's responses, ensuring that it incorporates relevant external knowledge into its generated answers.

\section{Related Paper}
\label{sec:Related-Paper}
\subsection{Relevance of LLM's in static vulnerability detection}
\label{subsec:Related-Paper/Relevance}
Several papers have already demonstrated the relevance of utilizing LLM's for static vulnerability detection, such as \textit{"The Emergence of Large Language Models in Static Analysis"} \cite{emergence2023} and \textit{"SkipAnalyzer"} \cite{skipanalyzer2023}.
\subsection{Privacy concerns in LSAST}
\label{subsec:Related-Paper/Privacy}
The paper \textit{"Large Language Model for Vulnerability Detection"} \cite{largemodel2023} emphasizes the importance of using locally hosted LLM's to address privacy concerns associated with third-party LLM providers. However, the paper has not benchmarked these locally hostable open-source LLM's.
\subsection{Importance of Finetuning in LSAST}
\label{subsec:Related-Paper/Finetuning}
Papers like \textit{"SecureFalcon"} \cite{securefalcon2023} and \textit{"Finetuning Large Language Models for Vulnerability Detection"} \cite{finetuning2024} have shown that fine-tuning LLM's can significantly enhance their vulnerability detection capabilities. These papers also recognize that the primary limitations of LLM's in vulnerability detection extend beyond just model accuracy.
\subsection{Knowledge retrieval in LSAST}
\label{subsec:Related-Paper/Knowledge}
The paper \textit{"LLM4Vuln"} \cite{llm4vuln2024} highlights the critical role of robust knowledge retrieval when utilizing LLM's for vulnerability scanning. It demonstrates LLM's capabilities in scanning smart contract code and emphasizes that effective knowledge retrieval is crucial for building an LLM-based SAST scanner. However, the paper mainly explores various components of such a scanner, with the knowledge retrieval system playing a relatively minor role in their research.

Our paper proposes not only combining traditional SAST scanner results with LLM's but also enhancing the knowledge retrieval aspect in multiple ways. We introduce new strategies for retrieving relevant vulnerability knowledge to support the LLM during the analysis phase.

Another paper, \textit{"Vul-RAG"} \cite{vulrag2024}, implements a knowledge retrieval system using Retrieval-Augmented Generation (RAG). While similar to our approach in using code-based similarity search, it does not combine functionality-similarity search or assist conservative SAST scanners, opting instead for a standalone approach.
\subsection{Assisting conservative SAST scanners with LLM's}
\label{subsec:Related-Paper/Knowledge}
Research on assisting conservative SAST scanners with LLM's is sparse. \textit{"The Hitchhiker’s Guide to Program Analysis"} \cite{hitchhiker2023} recognizes the assistive role of LLM's in static analysis and presents an initial approach for achieving synergy with LLM's in static analysis. However, it focuses on a specific case study of UBI bugs (Use-Before- Initialization) in Linux kernels and does not leverage specific static analysis tools or support a wide range of programming languages as our approach does. In our work we evaluate LSAST on a wide range of programming languages including javascript, php, java and others.

Another relevant paper, \textit{"LLM-Assisted Static Analysis for Detecting Security Vulnerabilities"} \cite{llmstatic2023} leverages LLM's to complement traditional static analysis techniques, focusing on using CodeQL for detailed code querying and analysis. This approach does not directly integrate SAST scanner outputs into the LLM prompt or implement knowledge retrieval systems as described in LSAST, but it has the potential to be integrated with our research in the future.

\section{Methodology}
\label{sec:Methodology}
\subsection{Combining conservative SAST scanners with LLM's}
\label{subsec:Methodology/include-conservative-SAST}
The main aspect of LSAST is the combination of a traditional SAST scanner with a large language model (LLM). To implement LSAST, it is crucial to choose an appropriate SAST scanning tool. We selected the open-source SAST scanner \href{https://github.com/Bearer/bearer}{Bearer} due to its fast response times and its ability to output data in various formats, which is useful for integrating the results into the prompt for the LLM.

\begin{lstlisting}[language=python, caption=single Bearer scan finding (simplified)]
{
      "cwe_ids": [
          "78"
      ],
      "id": "javascript_lang_os_command_injection",
      "title": "Unsanitized user input in OS command",
      "description": "Executing operating system commands with unsanitized ...",
      "line_number": 39,
      "full_filename": "/Users/User/repos/dvna/core/appHandler.js",
      "source": {
          "start": 39,
          "end": 44,
      }
}
\end{lstlisting}

In \textit{Listing 1}, you can see an example of a simplified output from a single Bearer scan finding. This output includes crucial details, such as the type of vulnerability and the affected lines.

In \textit{Listing 2}, you can see how we formatted the output of a Bearer scan.

\begin{lstlisting}[language=python, caption=formated output of Bearer scan results]
[
		'CWE-943 (line 59-65)', 
		'CWE-943 (line 85-89)', 
		'CWE-943 (line 107-111)', 
		'CWE-943 (line 145-149)', 
		'CWE-611 (line 235)', 
		'CWE-78 (line 39-44)'
]
\end{lstlisting}

\begin{lstlisting}[language=python, caption=target code with line information added]
line 1: var db = require('../models')
line 2: 
line 3: module.exports.userSearch = function (req, res) {
line 4: 	var query = "SELECT name,id FROM Users WHERE login='" + req.body.login + "'";
line 5: 	db.sequelize.query(query, {
line 6: 		model: db.User
line 7: 	}).then(user => {
line 8: 		if (user.length) {
line 9: 			var output = {
...
\end{lstlisting}

These results can now be integrated into the prompt for the LLM. Additionally, we need to provide the LLM with the target code we want to analyze. After importing the code file as a string, we adjust each line of the code to include line number information.

In \textit{Listing 3}, you can observe the appearance of the target code string after the necessary adjustments. Including line number information in the code ensures that the SAST response, which also contains line number information, can be interpreted effectively by the LLM. Failure to include line numbers could result in the loss of the LLM's ability to utilize the SAST scanner results comprehensively, potentially leading to inaccuracies in the LLM's own findings regarding line-specific vulnerabilities.

\begin{lstlisting}[language=python, caption=LLM prompt with bearer results]
Code:
{target_code}

After scanning the code with a SAST Scanner we receive following vulnerabilities:
{bearer_result}

You are a very efficient vulnerability scanner.
Follow these tasks carefully:
- Evaluate whether the given code includes any more vulnerabilities other then the ones found by the SAST Scanner.
- Only output vulnerabilities that were not found by the SAST Scanner.
- There can be multiple vulnerabilities
- Only output vulnerabilites that are 100 percent certain to be present in the given code.
- If you dont find any additional vulnerabilties just answer: "no additional vulnerabilties found"
- If you find a vulnerability structure your answer like this:
"CWE-ID: <CWE-ID>
Reason: <reason why code is vulnerable>
line: <the lines that cause the vulnerability>
code-snippet: <the code-snippet that cause the vulnerability>"
\end{lstlisting}
\textit{Listing 4} presents the complete prompt, encompassing both the target code and the results from the Bearer scan. This integrated prompt is now ready for direct querying by the LLM.
\subsection{Gathering Vulnerability Reports}
\label{subsec:Methodology/gathering-reports}
To ensure our queries include the most current vulnerability information, we've implemented a Knowledge Retrieval System. This system aggregates vulnerability reports from the past year, offering insights into newly exploitable vulnerabilities that may not be covered in the LLM's training data.

To begin, we retrieve these reports and prepare them for a similarity search. We utilize the Hacktivity API from \cite{hackerone} for gathering vulnerability data. HackerOne, a leading Bug Bounty Platform, hosts programs where companies incentivize security researchers to uncover vulnerabilities in their systems. Some findings from these programs are openly disclosed on HackerOne and accessible via their API. A collection of vulnerability reports is invaluable for LLM-supported DAST or SAST scanners, particularly in scenarios involving both Black-Box (application-only access) and White-Box (full code access) approaches.

Our efforts have yielded 873 vulnerability reports in JSON format, all dated within the past year and containing software code details.

In \textit{Listing 5}, you can find a simplified example of a single vulnerability report. This report includes the title, severity rating (ranging from low to critical), CVE IDs, CWE ID and title, and most importantly, vulnerability information. The vulnerability information section provides a detailed description of the vulnerability along with relevant code examples.

\begin{lstlisting}[language=python, caption=HackerOne vulnerability report (simplified), float]
{
    "id": 2493548,
    "title": "Incorrect Type Conversion in interpreting IPv4-mapped IPv6 addresses and below `curl` results in indeterminate SSRF vulnerabilities.",
    "severity_rating": "critical",
    "cve_ids": [
        "CVE-2023-24329",
        "CVE-2024-22243"
    ],
    "cwe-id": "CWE-843",
    "cwe-title": "Type Confusion",
    "vulnerability_information": "## Summary:\nOctal Type Handling of Errors in IPv4 Mapped IPv6 Addresses in curl allows unauthenticated remote attackers to ....",
},
\end{lstlisting}

\subsection{Preparing Vulnerability Reports for Knowledge Retrieval Process}
\label{subsec:Methodology/Preparing-Vulnerability-Reports}
To deliver these vulnerability reports effectively to the LLM, we employ a Retrieval-augmented System, which enhances the LLM's internal representation of information. Given that supplying all 873 vulnerability reports could potentially overwhelm and distract the LLM, our approach focuses on providing only the most relevant reports for the code that has to be scanned.

Various methods can determine which reports are most relevant for the target code.

LLM4Vuln \cite{llm4vuln2024} introduced a method that involves a similarity search based on the functionality of the code.

The process begins with summarizing the functionality of each code snippet within the vulnerability reports and storing these summaries in a vector database. This task utilizes an LLM, and specific instructions can be found in \textit{Listing 6}. When scanning the target code (\textit{TC}) for vulnerabilities, we also summarize its functionality using the LLM. Subsequently, we conduct a similarity search in our vector database to find vulnerable code snippets (\textit{VC}) with similar functionalities. Once we have gathered multiple functionalities of vulnerable code, we correlate them back to the respective vulnerability reports using the sequence numbers stored in the vector database entries. This method allows us to extract and deliver the most relevant vulnerability reports to the LLM for analysis.

Given the weak correlation between code functionality and vulnerabilities, we have developed a new approach for conducting similarity searches.

A stronger correlation often emerges when examining code abstraction in relation to vulnerabilities. Abstraction, also known as code minimization, involves simplifying source code to its fundamental components essential for correct execution by compilers or interpreters. To implement a similarity search based on code abstraction rather than functionality, we begin by extracting the code from vulnerability reports and then applying code minification techniques. This process can be achieved using code minifiers or leveraging LLM capabilities, as detailed in \textit{Listing 7}. After minifying the snippets of vulnerable code, we store them in a vector database.

When we now scan the target code (\textit{TC}) for vulnerabilities, we first minimize the \textit{TC} and then search for vulnerable code (\textit{VC}) with similar minified representations. Subsequently, we aggregate multiple code abstractions and correlate them back to the original reports using sequence numbers, as previously explained. This focused approach ensures that we only gather vulnerability reports with structures highly resembling the target code, increasing the likelihood of identifying identical vulnerabilities that are present in both \textit{VC} and \textit{TC}. This method aligns closely with the methodology employed by conservative SAST scanners, which aim to identify vulnerable patterns within code.

\begin{lstlisting}[ caption=LLM prompt to get abstraction of target code, float]
{target_code}

Given the code snippet, please remove all contextual information and variable names, leaving only the syntactic structure intact. 
1. The task involves extracting the core syntactic structure of the code while removing all contextual information and variable names, leaving only the essential code pattern intact. 
2. This process requires meticulous attention to detail and a thorough understanding of the code's structure. 
3. Focus on identifying and preserving the fundamental syntax elements, such as loops, conditionals, function definitions, and method invocations, while disregarding any specific variable names, comments, or non-essential whitespace. 
4. The resulting isolated code pattern will serve as a basis for further analysis and comparison, aiding in the identification of similar code patterns across multiple codebases or vulnerability reports.
5. Use the format: Code-Pattern:[xxxx], placing the extracted Code Pattern inside the brackets. 
6. If there is no code only answer "Code-Pattern:[no code]"
\end{lstlisting}

\begin{lstlisting}[ caption=LLM prompt to get summarization of target code, float]
{target_code}

Follow these tasks:
1. Outline the operational purpose of the provided code using imperative language, such as 'Calculate bank account balance.' Your explanation should be brief, no more than one paragraph and within 40-50 words.
2. Ensure your description focuses solely on the functionality or business logic without referencing specific variables, functions, or expressions.
\end{lstlisting}

\subsection{Knowledge Retrieval Methods}
\label{subsec:Methodology/Knowledge-Retrieval-Methods}
In the following figures you can observe the various methods available for constructing the knowledge retrieval process through the combination of the functionality approach, abstraction approach and conservative SAST scanner results.
\begin{description}
    \item[TC:] Target Code
    \item[SR:] SAST-Scanner Result
    \item[VR:] Vulnerability Reports
    \item[FVR:] Functionality-similar Vulnerability Reports
    \item[CVR:] Code-Abstraction-similar Vulnerability Reports
    \item[k:] Value of top $k$ in the similarity search
\end{description}

\subsubsection{Own-Knowledge}
\label{subsubsec:Methodology/Knowledge-Retrieval-Methods/Own-Knowledge-RAW}

\begin{figure}[t]
\begin{center}
\includegraphics[scale=0.7]{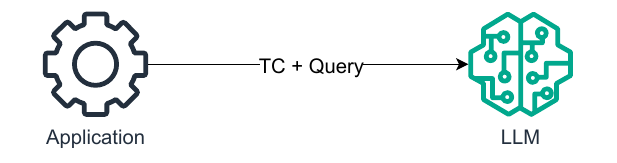}
\caption{Own-Knowledge Approach RAW}
\end{center}
\end{figure}

\begin{equation}
\text{llm-input} = \text{TC}
\end{equation}
This method is the most basic of all. Here we just simply query the LLM directly by including \textit{TC} in the scanning prompt, without using the knowledge retrieval system or bearer results.

\subsubsection{Own-Knowledge-LSAST (our new approach)}
\label{subsubsec:Methodology/Knowledge-Retrieval-Methods/Own-Knowledge-LSAST}

\begin{figure}[t]
\begin{center}
\includegraphics[scale=0.7]{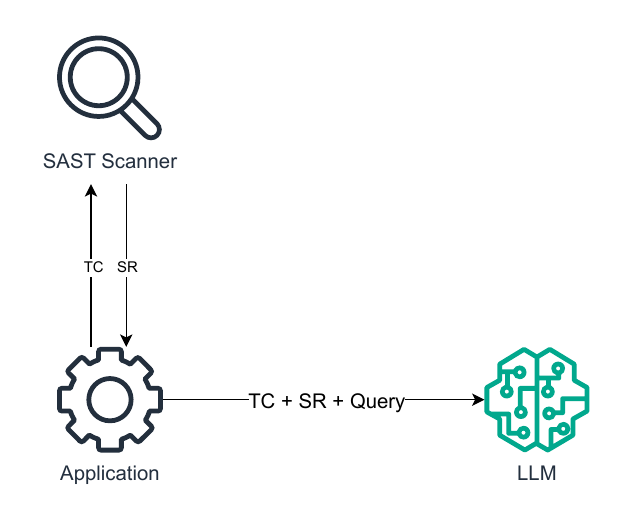}
\caption{Own-Knowledge Approach LSAST}
\end{center}
\end{figure}

\begin{equation}
\text{llm-input} = \text{TC} + \text{SR}
\end{equation}
This method enhances the Own-Knowledge Approach with the bearer scanner results.
We first run a bearer scan and then provide the results to the LLM, with the target code to analyze.

\subsubsection{Functionality}
\label{subsubsec:Methodology/Knowledge-Retrieval-Methods/Functionality}

\begin{figure}[t]
\begin{center}
\includegraphics[scale=0.7]{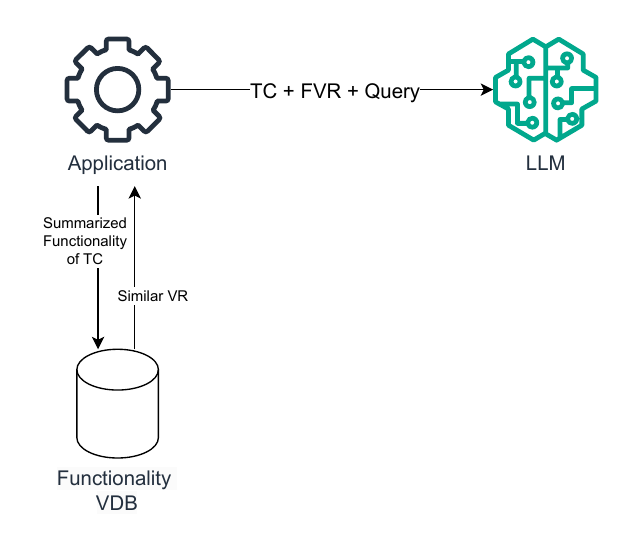}
\caption{Functionality Approach RAW}
\end{center}
\end{figure}

\begin{equation}
\text{llm-input} = \text{TC} + \text{FVR}
\end{equation}
Here we use the LLM to summarize the functionality of the \textit{TC}. We use the functionality of the \textit{TC} to make a similarity search inside the vector database that contains all the functionalities of the vulnerable code snippets. From the similarity search we gather the most similar \textit{VC} snippets, that are then used to gather the correlated \textit{VR's}. The \textit{VR's} from the similarity search are then provided to the LLM with the target code to analyze.

\subsubsection{Functionality-LSAST (our new approach)}
\label{subsubsec:Methodology/Knowledge-Retrieval-Methods/Functionality-LSAST}

\begin{figure}[t]
\begin{center}
\includegraphics[scale=0.7]{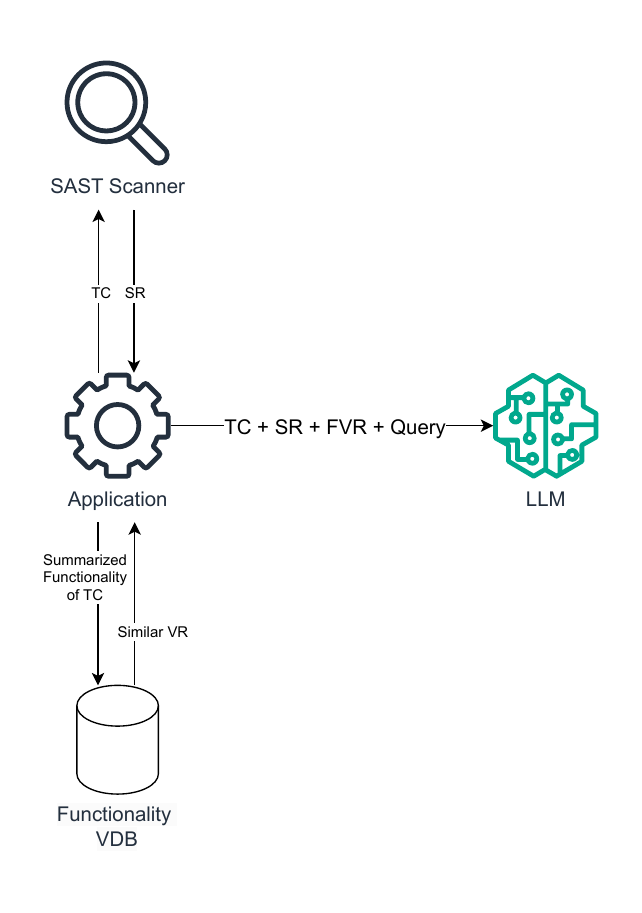}
\caption{Functionality Approach LSAST}
\end{center}
\end{figure}

\begin{equation}
\text{llm-input} = \text{TC} + \text{FVR} + \text{SR}
\end{equation}
Same as the functionality approach, but we also provide bearer scanner results to the LLM. 

\subsubsection{Abstraction (our new approach)}
\label{subsubsec:Methodology/Knowledge-Retrieval-Methods/Abstraction}

\begin{figure}[t]
\begin{center}
\includegraphics[scale=0.7]{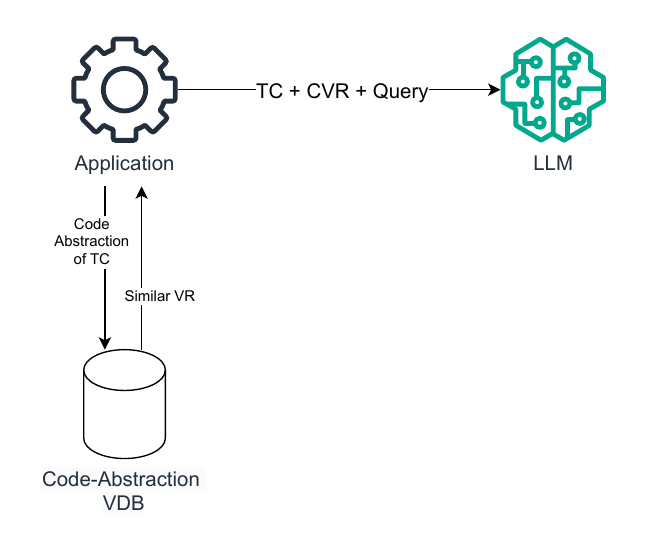}
\caption{Abstraction Approach RAW}
\end{center}
\end{figure}

\begin{equation}
\text{llm-input} = \text{TC} + \text{CVR}
\end{equation}
We use the LLM to abstract the \textit{TC}. We use the abstraction of the \textit{TC} to make a similarity search inside the vector database that contains all the code abstractions of the vulnerable code snippets. From the similarity search we gather the most similar \textit{VC} snippets, that are then used to gather the correlated \textit{VR's}. The \textit{VR's} from the similarity search are then provided to the LLM with the target code to analyze.

\subsubsection{Abstraction-LSAST (our new approach)}
\label{subsubsec:Methodology/Knowledge-Retrieval-Methods/Abstraction-LSAST}

\begin{figure}[t]
\begin{center}
\includegraphics[scale=0.7]{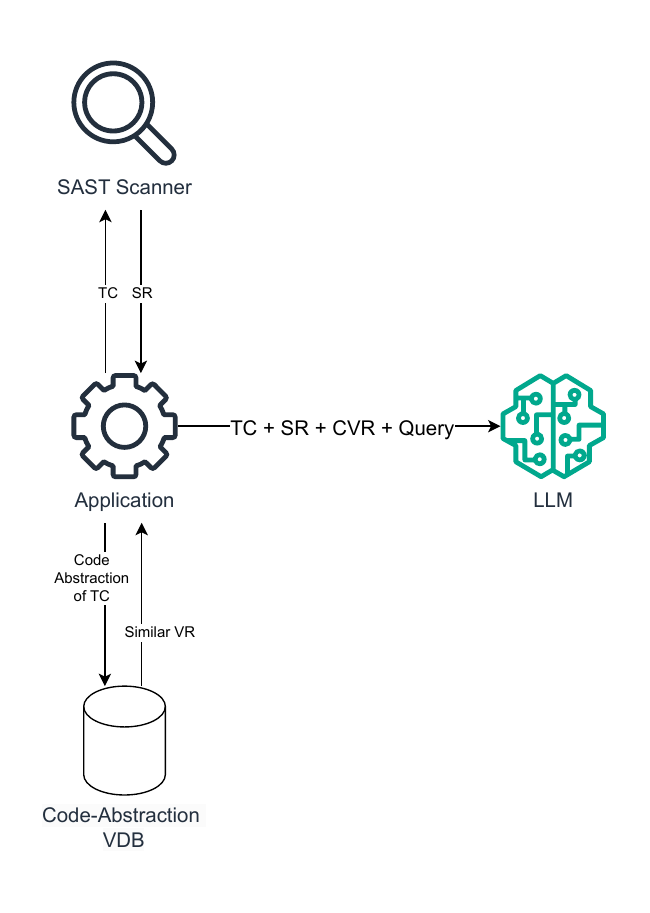}
\caption{Abstraction Approach LSAST}
\end{center}
\end{figure}

\begin{equation}
\text{llm-input} = \text{TC} + \text{CVR} + \text{SR}
\end{equation}
Same as the abstraction approach, but we also provide bearer scanner results to the LLM.

\subsubsection{Combined (our new approach)}
\label{subsubsec:Methodology/Knowledge-Retrieval-Methods/Combined}

\begin{figure}[t]
\begin{center}
\includegraphics[scale=0.5]{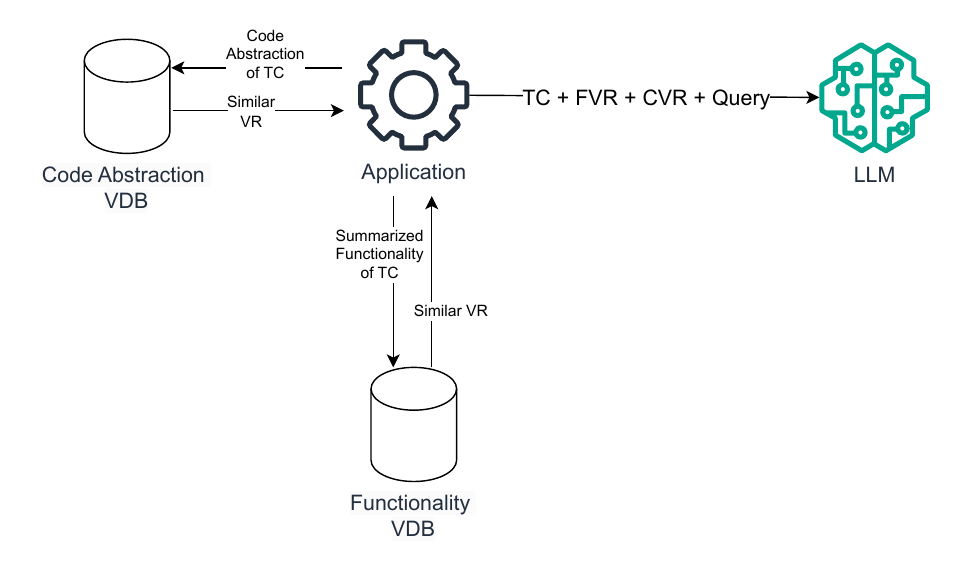}
\caption{Combined Approach RAW}
\end{center}
\end{figure}

\begin{equation}
\text{llm-input} = \text{TC} + \text{CVR} + \text{FVR}
\end{equation}
This approach combines Functionality and Pattern similarity search. k \textit{VR's} from the Functionality similarity search and k \textit{VR's} from the Pattern similarity search are gathered. These vulnerability reports are then provided to the LLM.

\subsubsection{Combined-LAST (our new approach)}
\label{subsubsec:Methodology/Knowledge-Retrieval-Methods/Combined-LSAST}

\begin{figure}[t]
\begin{center}
\includegraphics[scale=0.5]{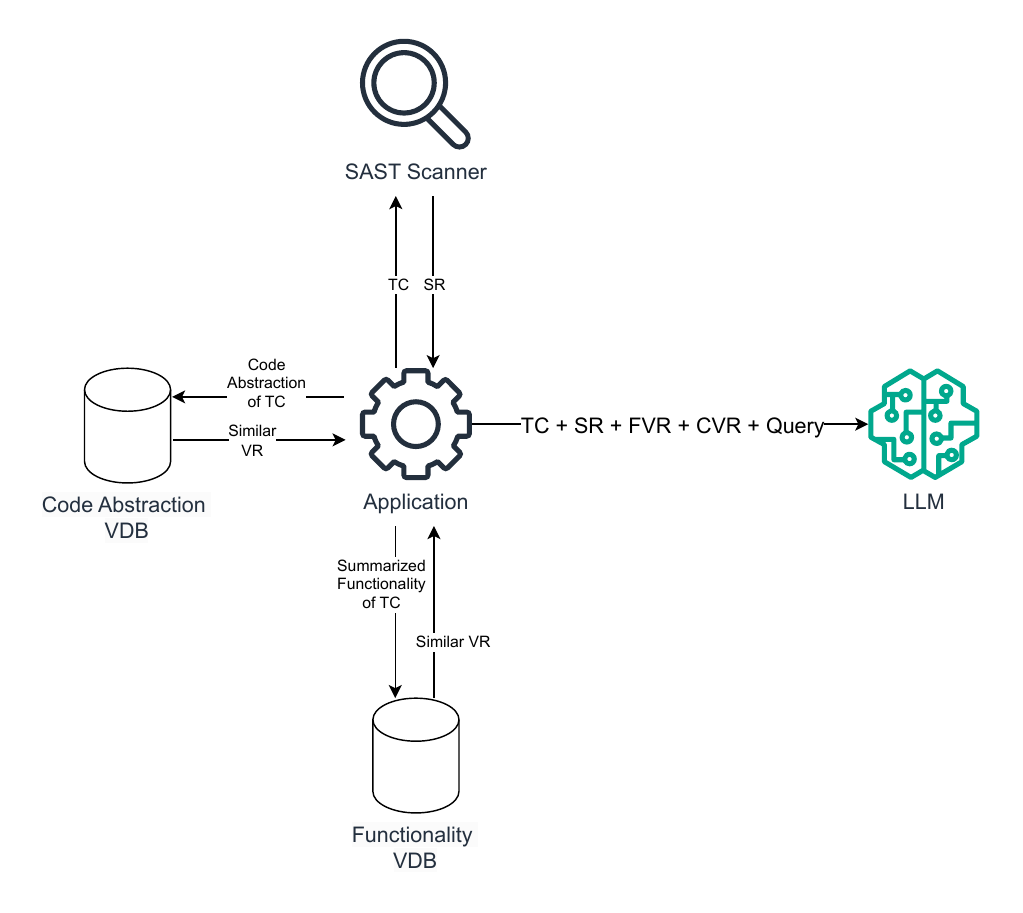}
\caption{Combined Approach LSAST}
\end{center}
\end{figure}

\begin{equation}
\text{llm-input} = \text{TC} + \text{CVR} + \text{FVR} + \text{SR}
\end{equation}
Inspired by \cite{jiang2024guidelight}, which also incorporates industrial solutions as guidance, we also provide bearer scanner results to the LLM, and the other settings remain the same as the combined approach.

\section{Evaluation}
\label{sec:Evaluation}
Since previous work has already shown that LLM's can be utilized for vulnerability scanning and it is very time-consuming to evaluate all approaches, we have decided to structure our approach in the most efficient way. First, for all evaluations, we are using Llama-3-70B, as it is a high-performing mid-sized model that can be hosted without needing very large cloud infrastructures or sending the code to third-party LLM providers. Also, we have first evaluated which of the RAG approaches is most efficient in a small-scale evaluation before attempting the big evaluation between the raw models and the best-performing RAG approach.

\subsection{Abstraction vs Functionality vs Combined}
\label{subsec:Evaluation/compare-rag}

\begin{table}[h]
\centering
\begin{tabular}{|l|c|c|c|}
\hline
\textbf{Measure} & \textbf{Abstraction} & \textbf{Functionality} & \textbf{Combined} \\
\hline
TP-Rate & 23.81\% & 14.29\% & 28.57\% \\
FP-Rate & 100\% & 100\% & 100\% \\
Accuracy & 19.23\% & 10\% & 24\% \\
Precision & 50\% & 25\% & 60\% \\
\textbf{F1-Score} & \textbf{32.26\%} & \textbf{18.18\%} & \textbf{38.71\%} \\
\hline
\end{tabular}
\caption{Comparison of Abstraction, Functionality, and Combined Approaches}
\label{table:comparison1}
\end{table}
\noindent We started by evaluating the \hyperref[subsubsec:Methodology/Knowledge-Retrieval-Methods/Abstraction]{Abstraction Approach (4.4.5)}, \hyperref[subsubsec:Methodology/Knowledge-Retrieval-Methods/Functionality]{Functionality Approach (4.4.3)}, and the \hyperref[subsubsec:Methodology/Knowledge-Retrieval-Methods/Combined]{Combined Approach (4.4.7)} without feeding the bearer input into the prompt. We evaluated our approaches using four well-known open-source security testing projects: 
\begin{enumerate}
    \item \textbf{DVWA (Damn Vulnerable Web Application) \cite{dvwa}}: A PHP/MySQL web application intentionally designed with vulnerabilities to help security professionals test their tools and hone their vulnerability detection skills.
    \item \textbf{DVNA (Damn Vulnerable NodeJS Application) \cite{dvna}}: A Node.js web application that intentionally incorporates OWASP Top 10 Vulnerabilities to demonstrate how these vulnerabilities can be identified and mitigated.
    \item \textbf{OWASP Juice Shop \cite{juice}}: 
    A modern JavaScript-based application (using Node.js, Express, and Angular) intentionally built with vulnerabilities. It serves as a platform for security training and testing of security tools.
    \item \textbf{WebGoat \cite{goat}}: A Java-based web application deliberately including vulnerabilities, designed for security professionals to explore and understand common Java-specific security issues.
\end{enumerate}
The projects include different programming languages like Java, JavaScript, TypeScript, and PHP and are built to include a wide range of different vulnerabilities. Across all 4 projects, using our 3 scanner approaches, we have scanned a total of 10 files that include more than 15 different kinds of vulnerabilities. 

For our evaluation metrics, we used standard measures in vulnerability detection:
\begin{itemize}
    \item \textbf{True Positive Rate (TP-Rate)}: The proportion of reported vulnerabilities that were actual vulnerabilities.
    \item \textbf{False Positive Rate (FP-Rate)}: The proportion of vulnerabilities that were overlooked.
    \item \textbf{Accuracy}: The ratio of correctly predicted instances to the total number of predictions.
    \item \textbf{Precision}: The proportion of predicted vulnerabilities that were actually vulnerabilities.
    \item \textbf{F1-Score}: A single score that represents overall performance.
\end{itemize}
As ground truth, we have decided to use the Bearer results, challenging the 3 approaches to find as many vulnerabilities as the static scanner finds. This evaluation is to find out which of the 3 approaches covers most of the underlying vulnerabilities. 

In \textit{Table 1}, you can see that the Abstraction Approach has outperformed the Functionality Approach, but the Combined Approach has outperformed both the Abstraction Approach and the Functionality Approach. The 100\% FP-Rate is achieved since all files include at least 1 vulnerability, and therefore a scan never had the opportunity to label a file as false positive.

\subsubsection*{Interpretation}
As explained earlier, there is a higher correlation between code abstraction and underlying vulnerabilities compared to functionality and underlying vulnerabilities. This is because functionalities can differ greatly but still have the same vulnerabilities, and vice versa. Conservative scanners also scan for underlying code patterns or syntactic features. Therefore, we can see that our new Abstraction and Combined approaches work better than the Functionality approach, which was introduced in LLM4Vuln \cite{llm4vuln2024}.

The effectiveness of the Combined Approach implies that the additional vulnerability reports it gathers, relative to the Pattern Approach, contribute valuable information to the LLM, even if they lack high similarity in code abstraction.

The overall weak performance can be explained by examining the reports that were passed to the LLM depending on the chosen approach. We have looked at the vulnerability reports that were passed to the LLM after the selected similarity search approach and found that most of the reports were not related to the CWE's that were found by Bearer.
\begin{table}[H]
\centering
\begin{tabular}{|l|c|c|}
\hline
\textbf{Measure} & \textbf{Abstraction} & \textbf{Functionality} \\
\hline
TP-Rate & 33.33\% & 16.67\% \\
FP-Rate & 100\% & 100\% \\
Accuracy & 12.50\% & 6.90\% \\
Precision & 16.67\% & 10.53\% \\
\textbf{F1-Score} & \textbf{22.22\%} & \textbf{12.90\%} \\
\hline
\end{tabular}
\caption{Comparison of similarity search for Abstraction and Functionality Approaches}
\label{table:comparison2}
\end{table}\noindent
In \textit{Table 2}, you can see the results of the Similarity Search Performance. The TP-Rate can be interpreted as how many of the vulnerabilities that we passed to the LLM were also found by Bearer. In other words, around 67\% of the reports that we passed to the LLM, with the abstraction approach, were related to other vulnerabilities than the ones found by Bearer.
\subsubsection*{Interpretation}
We distract the LLM from finding the vulnerabilities that were identified by Bearer, by passing large numbers of vulnerabilities that were not found by Bearer to the LLM. 
\subsection{RAW vs LSAST vs LSAST-Combined}
\label{subsec:Evaluation/Compare-LLM}
 We have evaluated that the combined approach performs the best out of the three knowledge retrieval approaches. Now we want to compare the combined approach with the approaches that do not use a knowledge retrieval system. For this, we have again scanned the open-source projects DVWA, DVNA, Juice Shop, and WebGoat, but this time using the \hyperref[subsubsec:Methodology/Knowledge-Retrieval-Methods/Combined-LSAST]{combined LSAST-Approach (4.4.8)} (with knowledge retrieval and bearer results), the \hyperref[subsubsec:Methodology/Knowledge-Retrieval-Methods/Own-Knowledge-LSAST]{raw LSAST-Approach (4.4.2)} (without knowledge retrieval, but with bearer results), and the  \hyperref[subsubsec:Methodology/Knowledge-Retrieval-Methods/Own-Knowledge-RAW]{raw Approach (4.4.1)} (without knowledge retrieval and without bearer results). Since we now pass the Bearer results directly to the LLM for the combined LSAST-Approach and the raw LSAST-Approach, we want to analyze how many true vulnerabilities the scanners can find besides the ones Bearer already finds. To determine if the found vulnerabilities are true positives or false positives, we have gone through all the vulnerabilities detected by the scanners and labeled them as:
\begin{itemize}
    \item \textbf{True Positive}: If the predicted vulnerability is labeled by us as a real vulnerability that has not been found by Bearer.
    \item \textbf{False Positive}: If the predicted vulnerability is not a real vulnerability that can be exploited or is a hallucination.
    \item \textbf{False Negative}: If one of the other approaches has found a real vulnerability but they have not been found by this approach.
\end{itemize}
\noindent
We have labeled all vulnerabilities as duplicate if Bearer already detected the vulnerability and have not included them in the evaluation.

\begin{table}[h]
\centering
\begin{tabular}{|l|c|c|c|}
\hline
\textbf{Measure} & \textbf{raw} & \textbf{raw-LSAST} & \textbf{Combined-LSAST} \\
\hline
TP-Rate & 17.91\% & 68.89\% & 35.71\% \\
FP-Rate & 100\% & 100\% & 100\% \\
Accuracy & 24\% & 62\% & 32.61\% \\
Precision & 63.16\% & 86.11\% & 78.95\% \\
\textbf{F1-Score} & \textbf{38.71\%} & \textbf{76.54\%} & \textbf{49.18\%} \\
\hline
\end{tabular}
\caption{Comparison of Measures}
\label{tab:comparison_measures}
\end{table}
\noindent The results in \textit{Table 3} show that passing the Bearer findings to the LLM significantly increases the capabilities of LLM's in finding vulnerabilities that have not been found by a static scanner. Meanwhile, the Combined LSAST-Approach is not as high-performing as the raw LSAST-Approach, and the raw LLM performs worst.
\subsubsection*{Interpretation}
Two major takeaways can be drawn from the results of our evaluation:
\begin{enumerate}
\item Passing the SAST scanner results to the LLM increases the performance of the LLM in finding new vulnerabilities that can't be found by the SAST scanners.
\item A knowledge retrieval structure for delivering similar vulnerabilities to the LLM can be counterproductive when not carefully crafted.
\end{enumerate}
\subsubsection*{Explaination}
\begin{enumerate}
\item Passing the SAST scanner results to the LLM increases the performance of the LLM in finding new vulnerabilities that can't be found by the SAST scanners.
\end{enumerate}
Our work has proven a very important learning for using LLM's in static vulnerability code analysis. By informing the LLM of the vulnerabilities that are already present in the code, the LLM can focus on the analysis of other code segments and/or other vulnerabilities that may be present in the code.
\begin{enumerate}
\item[2] A knowledge retrieval structure for delivering similar vulnerabilities to the LLM can be counterproductive when not carefully crafted.
\end{enumerate}
Our results have shown that using a RAG infrastructure for knowledge retrieval can be counterproductive for the LLM's vulnerability scanning capabilities. The vulnerability reports that are being passed to the LLM to guide it in the code analysis and deliver up-to-date vulnerabilities have to be highly relevant to the code segment that needs to be analyzed, or it can distract the LLM's focus and decrease its performance.

\subsection{Real-world test}
\label{subsec:Evaluation/real}
To demonstrate that LSAST is capable of identifying vulnerabilities missed by conventional SAST scanners in real-world scenarios, we scanned the Python package pyDust \cite{pyDust} for vulnerabilities and examined the findings in collaboration with the repository's authors. PyDust is a complex Python package that implements an entity-attribute-value data model, providing mechanisms for creating, manipulating, and serializing entities with dynamic attributes, along with features for change tracking and artifact management.

After scanning the repository, LSAST correctly identified 2 High-Risk, 1 Medium-Risk, and 3 Low-Risk vulnerabilities that conventional scanners failed to detect. These vulnerabilities could lead to serious security issues, such as code injection or remote code execution in the application if left unfixed.

In addition to the 5 true positives, the scanner also reported 2 false positives, highlighting the importance of double-checking scanner results.

Given that PyDust is a sophisticated piece of software and LSAST was still able to find vulnerabilities, we have demonstrated that LSAST can perform effectively in real-world environments.

\section{Conclusion}
\label{sec:Conclusion}

\subsection{Future Work}
\label{subsec:Conclusion/future}
To improve the knowledge retrieval system, multiple steps must be taken to bring the performance to a state where the LLM benefits from the vulnerability reports without getting distracted:
\begin{itemize}
    \item A new and improved source of vulnerabilities should be introduced. This source should include the vulnerable code, the CWE-ID, and a description of the vulnerability. Our current source, HackerOne, had the problem of mixing code and description, which led us to use LLM's to extract the code from the description. This resulted in low-quality vulnerable code snippets.
    \item For the abstraction approach, we have used LLM's to transform the code into a code abstraction. Instead, algorithms should be used to ensure consistency.
\end{itemize}
Besides improvements in the knowledge retrieval system, other enhancements and research areas in LSAST can be explored:
\begin{itemize}
    \item Chain of thought reasoning should be applied.
    \item Research should be conducted on how to scan an entire repository without needing to chunk the request and while preserving context, to extend the LLM's understanding of the whole repository.
    \item Consistency of results should be investigated. One idea would be to apply ensemble methods to aggregate predictions from multiple scans with different hyperparameters, to find reliable answers, minimize false positives, and increase consistency.
\end{itemize}

\subsection{Impact of the paper}
\label{subsec:Conclusion/impact}
Our work has made the following contributions:
\begin{itemize}
    \item First, we have significantly enhanced LLM's capabilities in scanning for vulnerabilities by passing static code scanning findings to an LLM, enabling it to find vulnerabilities that would not have been detected if relying on conservative methods alone. By doing this, we have demonstrated the synergy that can be unlocked when combining these technologies.
    \item Second, we have addressed LLM's limitation in gathering up-to-date vulnerability information and have improved current methods for addressing this issue by introducing new approaches for knowledge retrieval systems. Our combined and abstraction approaches outperform the knowledge retrieval systems presented in other papers. These approaches can now be used as new standards for building and improving knowledge retrieval systems for static vulnerability scanning.
    \item Third, we have shown that using a knowledge retrieval system can be counterproductive if not perfectly crafted. Reports that are passed to the LLM can distract its focus if they are not highly relevant.
    \item Fourth, by using open-source LLM's in our research, we have mitigated privacy and security risks and have demonstrated that everyone can make use of the capabilities of LLM's in vulnerability scanning without compromising in these areas.
    \item Fifth, we have demonstrated in a real-world scenario that LSAST scanners can detect vulnerabilities missed by traditional scanners, making them essential components of any thorough security scanning process moving forward.
\end{itemize}
Overall, we have made important findings in the area of LLM-supported static code analysis. We have demonstrated the potential for synergy between traditional static analysis tools and LLM's and have shown how to utilize such synergy. 

While there is still much research to be done in this field, we can confidently say that LLM's can play a crucial part of the static code analysis process, and we expect to see significant improvements in the near future.

\bibliographystyle{plain}
\bibliography{main}

\begin{thebibliography}{10}

\bibitem{achiam2023gpt}
Josh Achiam, Steven Adler, Sandhini Agarwal, Lama Ahmad, Ilge Akkaya, Florencia~Leoni Aleman, Diogo Almeida, Janko Altenschmidt, Sam Altman, Shyamal Anadkat, et~al.
\newblock Gpt-4 technical report.
\newblock {\em arXiv preprint arXiv:2303.08774}, 2023.

\bibitem{dvna}
Appsecco, 2024.

\bibitem{forbes2021}
Chuck Brooks.
\newblock More alarming cybersecurity stats for 2021.
\newblock {\em Forbes}, 2021.
\newblock Accessed: 2024-06-19.

\bibitem{vulrag2024}
Xueying Du, Geng Zheng, Kaixin Wang, Jiayi Feng, Wentai Deng, Mingwei Liu, Bihuan Chen, Xin Peng, Tao Ma, and Yiling Lou.
\newblock Vul-rag: Enhancing llm-based vulnerability detection via knowledge-level rag.
\newblock {\em arXiv preprint arXiv:2401.17011}, 2024.
\newblock Accessed: 2024-06-24.

\bibitem{securefalcon2023}
Mohamed~Amine Ferrag, Ammar Battah, Norbert Tihanyi, Ridhi Jain, Diana Maimu¸t, Fatima Alwahedi, Thierry Lestable, Narinderjit~Singh Thandi, Abdechakour Mechri, Merouane Debbah, and Lucas~C. Cordeiro.
\newblock Securefalcon: Are we there yet in automated software vulnerability detection with llms?
\newblock {\em IEEE Transactions on Software Engineering}, 2023.
\newblock Accessed: 2024-06-24.

\bibitem{statista2023}
Anna Fleck.
\newblock Expected cost of cybercrime until 2027.
\newblock {\em Statista}, 2023.
\newblock Accessed: 2024-06-19.

\bibitem{hackerone}
HackerOne, 2024.

\bibitem{pyDust}
Zsolt Horvath, 2024.

\bibitem{jiang2024guidelight}
Haoyuan Jiang, Xuantang Xiong, Ziyue Li, Hangyu Mao, Guanghu Sui, Jingqing Ruan, Yuheng Cheng, Hua Wei, Wolfgang Ketter, and Rui Zhao.
\newblock Guidelight: "industrial solution" guidance for more practical traffic signal control agents.
\newblock {\em arXiv preprint arXiv:2407.10811}, 2024.

\bibitem{kong2024tptu}
Yilun Kong, Jingqing Ruan, Yihong Chen, Bin Zhang, Tianpeng Bao, Shiwei Shi, Guoqing Du, Xiaoru Hu, Hangyu Mao, Ziyue Li, et~al.
\newblock Tptu-v2: Boosting task planning and tool usage of large language model-based agents in real-world systems.
\newblock In {\em ICLR 2024: The Twelfth International Conference on Learning Representations Workshop on LLM Agents}, 2024.

\bibitem{hitchhiker2023}
Haonan Li, Yu~Hao, Yizhuo Zhai, and Zhiyun Qian.
\newblock The hitchhiker’s guide to program analysis: A journey with large language models.
\newblock In {\em Proceedings of the 2023 International Conference on Software Engineering}, 2023.
\newblock Accessed: 2024-06-24.

\bibitem{li2024pet}
Zhishuai Li, Xiang Wang, Jingjing Zhao, Sun Yang, Guoqing Du, Xiaoru Hu, Bin Zhang, Yuxiao Ye, Ziyue Li, Rui Zhao, et~al.
\newblock Pet-sql: A prompt-enhanced two-stage text-to-sql framework with cross-consistency.
\newblock {\em arXiv preprint arXiv:2403.09732}, 2024.

\bibitem{llmstatic2023}
Ziyang Li, Saikat Dutta, and Mayur Naik.
\newblock Llm-assisted static analysis for detecting security vulnerabilities.
\newblock {\em arXiv preprint}, 2023.
\newblock Accessed: 2024-06-24.

\bibitem{liu2024timecma}
Chenxi Liu, Qianxiong Xu, Hao Miao, Sun Yang, Lingzheng Zhang, Cheng Long, Ziyue Li, and Rui Zhao.
\newblock Timecma: Towards llm-empowered time series forecasting via cross-modality alignment.
\newblock {\em arXiv preprint arXiv:2406.01638}, 2024.

\bibitem{liu2024spatial}
Chenxi Liu, Sun Yang, Qianxiong Xu, Zhishuai Li, Cheng Long, Ziyue Li, and Rui Zhao.
\newblock Spatial-temporal large language model for traffic prediction.
\newblock In {\em IEEE MDM 2024: The 25th IEEE International Conference on Mobile Data Management}, 2024.

\bibitem{skipanalyzer2023}
Mohammad~Mahdi Mohajer, Reem Aleithan, Nima~Shiri Harzevili, Moshi Wei, Alvine~Boaye Belle, Hung~Viet Pham, and Song Wang.
\newblock Skipanalyzer: A tool for static code analysis with large language models.
\newblock In {\em Proceedings of the 2023 IEEE/ACM 45th International Conference on Software Engineering}, 2023.
\newblock Accessed: 2024-06-24.

\bibitem{juice}
OWASP, 2024.

\bibitem{goat}
OWASP, 2024.

\bibitem{ruan2023tptu}
Jingqing Ruan, Yihong Chen, Bin Zhang, Zhiwei Xu, Tianpeng Bao, Guoqing Du, Shiwei Shi, Hangyu Mao, Ziyue Li, Xingyu Zeng, et~al.
\newblock Tptu: Task planning and tool usage of large language model-based ai agents.
\newblock In {\em NeurIPS 2023: 37th Conference on Neural Information Processing Systems (NeurIPS 2023) - Workshop on Foundation Models for Decision Making}, 2023.

\bibitem{finetuning2024}
Alexey Shestov, Rodion Levichev, Ravil Mussabayev, Evgeny Maslov, Anton Cheshkov, and Pavel Zadorozhny.
\newblock Finetuning large language models for vulnerability detection.
\newblock {\em arXiv preprint}, 2024.
\newblock Accessed: 2024-06-24.

\bibitem{emergence2023}
Ashwin~Prasad Shivarpatna~Venkatesh, Samkutty Sabu, Amir~M. Mir, Sofia Reis, and Eric Bodden.
\newblock The emergence of large language models in static analysis: A first look through micro-benchmarks.
\newblock In {\em Proceedings of the 2023 ACM SIGSOFT International Symposium on Software Testing and Analysis}, 2023.
\newblock Accessed: 2024-06-24.

\bibitem{sui2023reboost}
Guanghu Sui, Zhishuai Li, Ziyue Li, Sun Yang, Jingqing Ruan, Hangyu Mao, and Rui Zhao.
\newblock Reboost large language model-based text-to-sql, text-to-python, and text-to-function -- with real applications in traffic domain.
\newblock {\em arXiv preprint arXiv:2310.18752v2}, 2023.

\bibitem{llm4vuln2024}
Yuqiang Sun, Daoyuan Wu, Yue Xue, Han Liu, Wei Ma, Lyuye Zhang, Miaolei Shi, and Yang Liu.
\newblock Llm4vuln: A unified evaluation framework for decoupling and enhancing llms’ vulnerability reasoning.
\newblock {\em arXiv preprint}, 2024.
\newblock Accessed: 2024-06-24.

\bibitem{team2023gemini}
Gemini Team, Rohan Anil, Sebastian Borgeaud, Yonghui Wu, Jean-Baptiste Alayrac, Jiahui Yu, Radu Soricut, Johan Schalkwyk, Andrew~M Dai, Anja Hauth, et~al.
\newblock Gemini: a family of highly capable multimodal models.
\newblock {\em arXiv preprint arXiv:2312.11805}, 2023.

\bibitem{dvwa}
Robin Wood.
\newblock Damn vulnerable web application, 2024.

\bibitem{yang2024sql}
Sun Yang, Qiong Su, Zhishuai Li, Ziyue Li, Hangyu Mao, Chenxi Liu, and Rui Zhao.
\newblock Sql-to-schema enhances schema linking in text-to-sql.
\newblock In {\em DEXA 2024: The 35th Database and Expert Systems Applications Conferences and Workshops}, 2024.

\bibitem{ye2024survey}
Jiexia Ye, Weiqi Zhang, Ke~Yi, Yongzi Yu, Ziyue Li, Jia Li, and Fugee Tsung.
\newblock A survey of time series foundation models: Generalizing time series representation with large language mode.
\newblock {\em arXiv preprint arXiv:2405.02358}, 2024.

\bibitem{zhang2024controlling}
Bin Zhang, Hangyu Mao, Jingqing Ruan, Ying Wen, Yang Li, Shao Zhang, Zhiwei Xu, Dapeng Li, Ziyue Li, Rui Zhao, et~al.
\newblock Controlling large language model-based agents for large-scale decision-making: An actor-critic approach.
\newblock In {\em ICLR 2024: The Twelfth International Conference on Learning Representations Workshop on LLM Agents}, 2024.

\bibitem{zhang2024benchmarking}
Bin Zhang, Yuxiao Ye, Guoqing Du, Xiaoru Hu, Zhishuai Li, Sun Yang, Chi~Harold Liu, Rui Zhao, Ziyue Li, and Hangyu Mao.
\newblock Benchmarking the text-to-sql capability of large language models: A comprehensive evaluation.
\newblock {\em arXiv preprint arXiv:2403.02951}, 2024.

\bibitem{zhang2024dualtime}
Weiqi Zhang, Jiexia Ye, Ziyue Li, Jia Li, and Fugee Tsung.
\newblock Dualtime: A dual-adapter multimodal language model for time series representation.
\newblock {\em arXiv e-prints}, pages arXiv--2406, 2024.

\bibitem{largemodel2023}
Xin Zhou, Ting Zhang, and David Lo.
\newblock Large language model for vulnerability detection: Emerging results and future directions.
\newblock {\em ACM Computing Surveys}, 2023.
\newblock Accessed: 2024-06-24.

\end{thebibliography}

\end{document}